\documentclass[reprint, showpacs, superscriptaddress, aps, pre, twocolumn, 10pt]{revtex4-1}
\usepackage{graphicx}
\usepackage{color}
\usepackage{amssymb}
\usepackage{amsmath}
\usepackage{tabularx}
\usepackage{bm}
\usepackage{hyperref}
\usepackage{ltxtable}
\usepackage{natbib} 
\usepackage{longtable}
\usepackage{float}
\newcommand{\la}{\left\langle}
\newcommand{\ra}{\right\rangle}
\newcommand{\be}{\begin{equation}}
\newcommand{\ee}{\end{equation}}
\newcommand{\bse}{\begin{subequations}}
	\newcommand{\ese}{\end{subequations}}
\newcommand{\bea}{\begin{eqnarray}}
\newcommand{\eea}{\end{eqnarray}}
\newcommand{\ba}{\begin{array}}
	\newcommand{\ea}{\end{array}}

\usepackage[usenames,dvipsnames]{xcolor}
\hypersetup{
colorlinks,
citecolor=blue,
filecolor=blue,
linkcolor=blue,
urlcolor=blue}

\begin{document}

\title{ Equilibrium states of  Burgers and KdV equations }  
\author{Mahendra K. Verma}
\email{mkv@iitk.ac.in}
\affiliation{Department of Physics, Indian Institute of Technology Kanpur, Kanpur 208016, India}
\author{Soumyadeep Chatterjee}
\email{soumyade@iitk.ac.in}
\affiliation{Department of Physics, Indian Institute of Technology Kanpur, Kanpur 208016, India}
\author{Aryan Sharma}
\email{aryans@iitk.ac.in}
\affiliation{Department of Physics, Indian Institute of Technology Kanpur, Kanpur 208016, India}
\author{Ananya Mohapatra}
\email{amohapa2@ur.rochester.edu}
\affiliation{Department of Physics, Rochester University, Rochester, NY 14627-0171, USA}

\date{\today}

\begin{abstract}
We simulate KdV and dissipation-less Burgers equations using delta-correlated random noise as initial condition.  We observe that the energy fluxes of the two equations remain zero throughout, thus indicating their equilibrium nature.  We characterize the equilibrium states using Gaussian probability distribution for the real space field, and using Boltzmann distribution for  the modal energy.  We  show that the single soliton of the KdV equation too exhibits zero energy flux, hence it is in equilibrium.  We argue that the energy flux is a good measure for ascertaining whether a system is in equilibrium or not.
\end{abstract}
\maketitle

\section{Introduction} \label{sec:Introduction}

Thermodynamics systems, which are in equilibrium,  are well understood~\cite{Reif:book:StatMech}.  On the other hand, systems out of equilibrium are in general more complex, and they do not have a general framework so far.  Hydrodynamic and Burgers turbulence, crack propagation, and earthquakes are some of the prominent examples of nonequilibrium systems.  Interestingly, dissipation-less hydrodynamic and Burgers turbulence exhibit both nonequilibrium and equilibrium behaviour depending on the initial condition~\cite{Cichowlas:PRL2005,Ray:PRE2011,Verma:PTRSA2020}.  This is the subject of this paper.

Burgers equation is one of the well-studied equations of physics~\cite{Burgers:AAM1948,Bateman:MWR1915}. For smooth initial condition and small viscosity, Burgers equation admits shocks~\cite{Kida:JFM1979,Majda:PNAS2000,Bec:PR2007,Ray:PRE2011,Verma:PA2000}. The width of the shock region  increases with the increase of kinematic viscosity. 
However, dissipation-less Burgers equation exhibits different behaviour.    \citet{Majda:PNAS2000} studied the Galerkin-truncated dissipation-less Burgers equation with only a small number of modes and argued that the system is ergodic.  \citet{Ray:PRE2011} and \citet{Murugan:PRR2020} simulated the Galerkin-truncated Burgers equation with smooth initial condition (e.g., $ \sin(x) $) and observed small-scale oscillations, called ``tyger", superposed with shocks. The asymptotic solution of this system is dominated by tygers,  which are signatures of equilibrium behaviour.

\citet{Ray:PRE2011}'s  solution of the Burgers equation resembles \citet{Cichowlas:PRL2005}'s solution for the Euler equation.  \citet{Cichowlas:PRL2005} simulated Euler equation with Taylor-Green vortex as an initial condition and observed a mixed configuration whose  intermediate scales exhibit $ k^{-5/3} $ spectrum, while the  small scales exhibit equilibrium behaviour with $ k^2 $ spectrum.  However, the  asymptotic solution is dominated by $ k^2 $  spectrum, consistent with the predictions of \citet{Lee:QAM1952} and \citet{Kraichnan:JFM1973} on equilibrium states of Euler turbulence.

As we show in this paper, a convenient way to achieve equilibrium state of Burgers equation is to start the system with delta-correlated random initial condition.  We observe equipartition of energy among the Fourier modes in lines with the results of \citet{Majda:PNAS2000}, who  worked with small number of Fourier modes. The real-space velocity field exhibits Gaussian probability distribution function (PDF).  For this case, we show that the energy flux vanishes  throughout the evolution of the system.  

In this paper, we also study the equilibrium states of KdV equation, which is an important equation of physics.  The KdV equation has no dissipation, and hence it is a conservative system.  This equation exhibits a single soliton, as well as multiple solitons.   In this paper, we show that the single soliton solution is in equilibrium.

It has been argued that the KdV equation is a continuum representation of Fermi-Pasta-Ulam-Tsingou model~\cite{FermiPasta:1955,Benettin:JSP2008}.   
Note that  \citet{FermiPasta:1955} constructed a spring-mass model with nonlinear coupling and explored whether the system approaches its equilibrium state.  \citet{Benettin:JSP2008} studied the effects of initial condition on the approach to equilibrium.   They argued that for random phases of the  initially-excited modes, the energy per mode is the dominant parameter.  However, when we choose coherent phases for the initial modes,  the total energy is the dominant parameter. 

In this paper, we employ energy flux as a marker for testing equilibrium nature of the  Burgers and KdV equations.   The energy flux vanishes for the systems under equilibrium, while it is nonzero for nonequilibrium ones. Note that zero energy flux implies \textit{detailed balance of energy transfer} among the Fourier modes. The delta-correlated random solutions of   dissipation-less Burgers, KdV, and Euler equations are under equilibrium and they yield zero energy flux~\cite{Kraichnan:JFM1973,Lee:QAM1952,Verma:PTRSA2020}.   However,   the dissipative Burgers and hydrodynamic turbulence with large-scale forcing exhibit constant nonzero energy flux with  power-law ($ k^{-2} $ and $ k^{-5/3} $ respectively) energy spectra~\cite{Kida:JFM1979,Majda:PNAS2000,Verma:PA2000,Kolmogorov:DANS1941Dissipation,Kolmogorov:DANS1941Structure,Frisch:book}.  Note, however, that for smooth initial condition, the dissipation-less Burgers and Euler equations exhibit nonzero energy fluxes in the intermediate scales, and zero energy flux at small scales~\cite{Cichowlas:PRL2005,Ray:PRE2011,DiLeoni:PRF2018}.  We remark that the energy flux has not been used widely for characterizing nonequilibrium/equilibrium systems~\cite{Verma:book:ET,Verma:JPA2022}.

The outline of the paper is as follows.  In Sec.~\ref{sec:Theory} we describe the energy fluxes in KdV and Burgers Turbulence.  Section~\ref{sec:Numerical_Method} has a brief discussion of the numerical method, while Sec.~\ref{sec:Results} contains the results for the noisy KdV and dissipation-less Burgers turbulence.  In Sec.~\ref{sec:Soliton} we show that the soliton solution of KdV equation is in equilibrium.  We conclude in Sec.~\ref{sec:conclusions}. 

\section{Energy fluxes in KdV and Burgers Turbulence} \label{sec:Theory}

The Burgers equation~\cite{Bateman:MWR1915, Burgers:AAM1948}   and KdV equation~\cite{Drazin:book:Solitons}
are as follow:
\bea
\partial_t u + u \partial_x u & = & \nu \partial_x^2 u, \label{eq:burgers} \\
\partial_t u + u \partial_x u & = & -K \partial_x^3 u,
\label{eq:kdV}
\eea
where $ \partial_t, \partial_x $ are partial derivatives with relative to $ t $ and $ x $ respectively, $ u $ is the field variable, and $\nu, K$ are constants.  For the Burgers equation, $ u $ and $ \nu $ are interpreted  as the velocity field and kinematic viscosity of the fluid, respectively.  In this paper, we focus on the dissipation-less Burgers equation for which $ \nu=0 $.  Note that the  KdV equation with nonzero $ K $ is dissipation-less, but it exhibits  dispersion.  Consequently, the total energy, $ \int dx (u^2/2) $, is conserved for both the equations.  Since the forms of two equations are quite similar, for constructing a general formalism, we rewrite them as the following  equation:
\be
\partial_t u + u \partial_x u  =  D(x).
\label{eq:real_space}
\ee
We choose $ D(x)=0 $ for the Burgers equation, and $ D(x) = - K \partial_x^3 u $ for the KdV equation.

 The dissipation-less Burgers equation exhibits shocks for smooth initial conditions; here, the nonlinear term, $ u\partial_x u $, steepens the $ u $ profile~\cite{Kida:JFM1979}.  In the KdV equation, shock formation is prevented by the diffusion term, $ - K \partial_x^3 u$.  Instead,  the KdV equation exhibits solitons in which  the steepening (due to nonlinearity) is balanced by  diffusion.  Note, however, that we will focus on the solutions of  the dissipation-less Burgers and KdV equations with delta-correlated random initial condition.

To decipher the equilibrium nature and energy transfers of the KdV and Burgers equations, it is convenient to work in Fourier space.  In Fourier space,  Eq.~(\ref{eq:real_space}) gets transformed	to
\be
\frac{d}{dt} \hat{u}(k) + i k \sum_{p} \hat{u}(q) \hat{u}(p) =  \hat{D}(k),
\label{eq:Fourier_space}
\ee
where $q=k-p $, and $ \hat{u}(k), \hat{D}(k)$ are the Fourier transforms of $ u(x) $ and $ D(x) $ respectively.  {  Note that $\hat{D}(k) = 0$  and $ -i k^3  \hat{u}(k)$ for the dissipation-less Burgers and KdV equations respectively.  In addition, we consider Fourier-truncated equations with  $ N $ Fourier modes. The \textit{modal energy} corresponding to $  \hat{u}(k) $ is defined as  $ E(k) = | \hat{u}(k)|^2/2 $.   According to Parseval's theorem, the total energy in real space equals the sum of modal energies, that is,
\be
E  = \frac{1}{L} \int dx \frac{1}{2} |u(x)|^2 = \sum_k E(k).
\label{eq:energy_conserve}
\ee

Using Eq.~(\ref{eq:Fourier_space}),  \citet{Majda:PNAS2000} derived that
\be
\sum_k \frac{\partial \dot{\hat{u}}(k)} {\partial {\hat{u}}(k)} = 0.
\ee
Consider the evolution of the above systems in the phase space formed by $ \{\hat{u}(k) \} $. In this space, the phase space volume is conserved during its evolution, which is the statement of ``Liouville's theorem"~\cite{Landau:book:StatMech,Majda:PNAS2000,Reichl:book:StatMech,Lesieur:book:Turbulence}.  These observations link KdV and dissipation-less Burgers equations to microcanonical ensemble.  With the above property, the invariant Gibbs measure for the modal energy is
\be
P(E(k))  = \beta \exp[-\beta E(k)],
\ee
where  $ \beta  $ is a constant. Note that 
\be
\int_0^\infty  P(E(k))   dE(k) = 1.
\ee
Also,
\be
\la E(k) \ra = \int_0^\infty  E(k) \beta \exp[-\beta E(k)] dE(k) = \frac{1}{\beta} = \frac{E}{N}.
\ee
That is, the total energy is equipartitioned among all the available Fourier modes, similar to the equipartition in equilibrium statistical mechanics. These observations indicate that the asymptotic states of dissipation-less Burgers and KdV equations are in equilibrium~\cite{Majda:PNAS2000}.}

We can arrive at the above conclusions using energy transfer arguments as well. Using Eq.~(\ref{eq:Fourier_space}) we derive
\be
\frac{d}{dt} E(k)  =  \sum_{p} \Im [k\hat{u}(q) \hat{u}(p) \hat{u}^*(k) ] = T(k),
\label{eq:fourier_space}
\ee
where $ \Im[.] $ represents the imaginary part of the argument, and $ T(k) $ represents the nonlinear energy transfer to Fourier mode $ \hat{u}(k)$ from all other modes. Note that the diffusive term of the KdV equation yields zero contribution to the energy equation.  Hence, the energy equations for both Burgers and KdV equations are the same.

\textit{Energy flux} is a useful quantity in turbulence.   We define $ \Pi(\bar{k}) $ (for $ \bar{k}>0 $) as the energy leaving the wavenumber region $ (-\bar{k},\bar{k}) $.  In terms of $ T(k) $, the energy flux is defined as~\cite{Verma:PA2000,Verma:PR2004,Verma:book:ET,Verma:JPA2022}
\be
\Pi(\bar{k}) = -\sum_{|k'|=0}^{\bar{k}} T(k').
\label{eq:Tk_Pik}
\ee
Under a steady state,  the average  rate of change of energy is zero, i.e., $ d\la E(k) \ra/dt =0 $.  Hence, using Eq.~(\ref{eq:fourier_space}), we deduce that for all $ k $'s, 
 \be
\la T(k) \ra = 0.  
\label{eq:zero_Tk}
\ee
Therefore, using Eqs.~(\ref{eq:Tk_Pik},\ref{eq:zero_Tk}),  we conclude that
 \be
 \la \Pi(\bar{k})  \ra = 0.
\label{eq:zero_flux}
\ee
Thus, the energy fluxes of the KdV and  dissipation-less Burgers equations are zero.

Vanishing of  $ \la T(k) \ra $ and $  \la \Pi(\bar{k})  \ra  $ implies that the system is under equilibrium, as in  equilibrium statistical mechanics.  The condition $ \la T(k) \ra = 0  $ for all $ k $'s indicates that the system respects \textit{detailed balance in energy transfers}, i.e., there is no net energy exchange from a  Fourier mode  to another Fourier mode. These statements are valid for both Burgers and KdV equations.

Fields under  equilibrium are in general delta-correlated, that is,
\be
\la u(x,t) u(x',t) \ra = \la u^2 \ra \delta(x-x').
\ee
The Fourier transform of the above equation yields~\cite{Verma:PTRSA2020}
\be
 \la  E(k) \ra =  \frac{1}{2} \la  |\hat{u}(k)|^2 \ra = \frac{E}{N},
 \label{eq:equilibrium_Ek}
 \ee
 where $ E $ is the total energy, which is distributed evenly or  equipartitioned among all (here, $ N $) the Fourier modes.    Thus,  the energy spectra of the KdV and dissipation-less Burgers equations are constant in $ k $, and   the phases of the Fourier modes are random.  The random phases of $ \hat{u}(k) $'s yield $ \la T(k) \ra = 0$ (see  Eq.~(\ref{eq:fourier_space})).  Thus, the formalisms based on Liouville's theorem and energy transfers lead to the same conclusion  regarding the equilibrium state.   
  Equations~(\ref{eq:fourier_space}, \ref{eq:zero_Tk}, \ref{eq:equilibrium_Ek}), which are related to the flux formalism, provide alternative framework for differentiating equilibrium and nonequilibrium systems.  

In this paper, we will take  delta-correlated random $ u $ as the initial condition and time advance the KdV and dissipation-less Burgers equations.  We show that  the above initial condition yields the aforementioned  equilibrium configuration   at all times. Note that most of the past works take smooth large-scale $ u $ (e.g., $ \sin(x) $) as initial condition that yields  energy transfers among various Fourier modes, leading to unsteady and far-from-equilibrium configurations.  For example, \citet{Ray:PRE2011}  simulated dissipationless Burgers turbulence ($ \nu=0 $)  using $ \sin(x) $  as an initial condition. They observed  coherent shocks in the background of random noise that are similar to the delta-correlated noise discussed earlier. They also reported that the noisy component  grows with time. Note that $ T(k) \ne 0 $  and $ \Pi(\bar{k})  \ne 0 $   for \citet{Ray:PRE2011}'s solutions because the system is unsteady.

There is an interesting anomaly. A soliton is a steady-state solution of KdV equation. Hence, following the above arguments, $ T(k) = 0 $ and $ \Pi(\bar{k})  = 0 $ for a soliton as well.  Note however that the soliton  is invariant in time. Hence, $\{ \hat{u}(k) \}$ is a point in the phase space, rather than covering the entire phase space, as is envisaged in ergodic hypothesis.  Thus, a soliton and the noisy $ u$'s of the Burgers and KdV equations have very different phase space structure even though all of them exhibit zero energy flux.

The above framework for Burgers and KdV equations is very similar to that of Euler turbulence.  \citet{Lee:QAM1952} and \citet{Kraichnan:JFM1973} showed that Euler turbulence  exhibits equilibrium behaviour with zero energy flux and constant modal energies for all the modes.  Clearly, the equilibrium configurations of  the KdV and dissipation-less Burgers equations discussed above are very similar to that of Euler turbulence.   \citet{Verma:PTRSA2020} and \citet{Verma:arxiv2020_equilibrium} simulated Euler turbulence using delta-correlated  random initial condition and  obtained the aforementioned equilibrium state.  Note that a smooth large-scale velocity field as an initial condition, as in \citet{Cichowlas:PRL2005}, yields nonzero energy flux as transients.  However, the system  reaches an equilibrium state asymptotically.

In the next section, we will outline the numerical method employed for  simulating dissipation-less Burgers and KdV turbulence.

\section{Numerical method} \label{sec:Numerical_Method}
We simulate KdV and Burgers equations (Eq.~(\ref{eq:real_space})) using pseudo-spectral  method~\cite{Boyd:book:Spectral}.  We prefer this method over others (e.g., finite-difference) due to its accuracy.  {  For the two equations, we employ
$G = 32768~(2^{15})$ grid points in a box of size $ 2\pi $. We employ the fourth-order Runge-Kutta method for time stepping with time-step $dt=10^{-6}$~\cite{Verma:book:Computing}.} 
In addition, we employ  $2/3$ rule for dealiasing.  Hence,  the number of active Fourier modes $ N = (2/3) G $. Consequently,   $k_\mathrm{max}=G/3$.

In this paper, we focus on the equilibrium solution, which is conveniently obtained using delta-correlated  random initial conditions.  For the same, we set 
\be
u(k,t=0) = |A|e^{i\phi(k)}  
\label{eq:random_uk}
\ee
with equal amplitude $ |A| $ for all the Fourier modes. For $ \phi (k)$'s, we choose different  random numbers from uniform distribution between $0$ and $2\pi$.  An inverse Fourier transform of the above $ \hat{u}(k) $ yields delta-correlated random field (also called \textit{white noise}).  In the next section, we show that the above initial condition yields $ u(x,t) $ that is as random as the initial condition, thus yielding equilibrium states at all times.

We choose $A$  such that the initial total energy {$E(t=0)=1$}. 
We simulate both the equations up to {$13$} non-dimensionalized time units; here, the unit time corresponds to  $ L/u_\mathrm{rms} $, where $ u_\mathrm{rms} $ is the rms value of $ u $.  
The parameter $ K $ of the KdV equation is taken to be {$10^{-10}$}.  

In the next section, we describe the numerical results of  KdV and dissipation-less Burgers equations.

\section{Results on the equilibrium states of  KdV and Burgers turbulence} \label{sec:Results}

In this section, we will compute the energy spectra and fluxes, probability distribution functions, and phase-space projections for the  KdV and dissipation-less Burgers  equations.

\subsection{Energy evolution; energy spectrum and flux}

\begin{figure}[h!]
	\includegraphics[width=\linewidth]{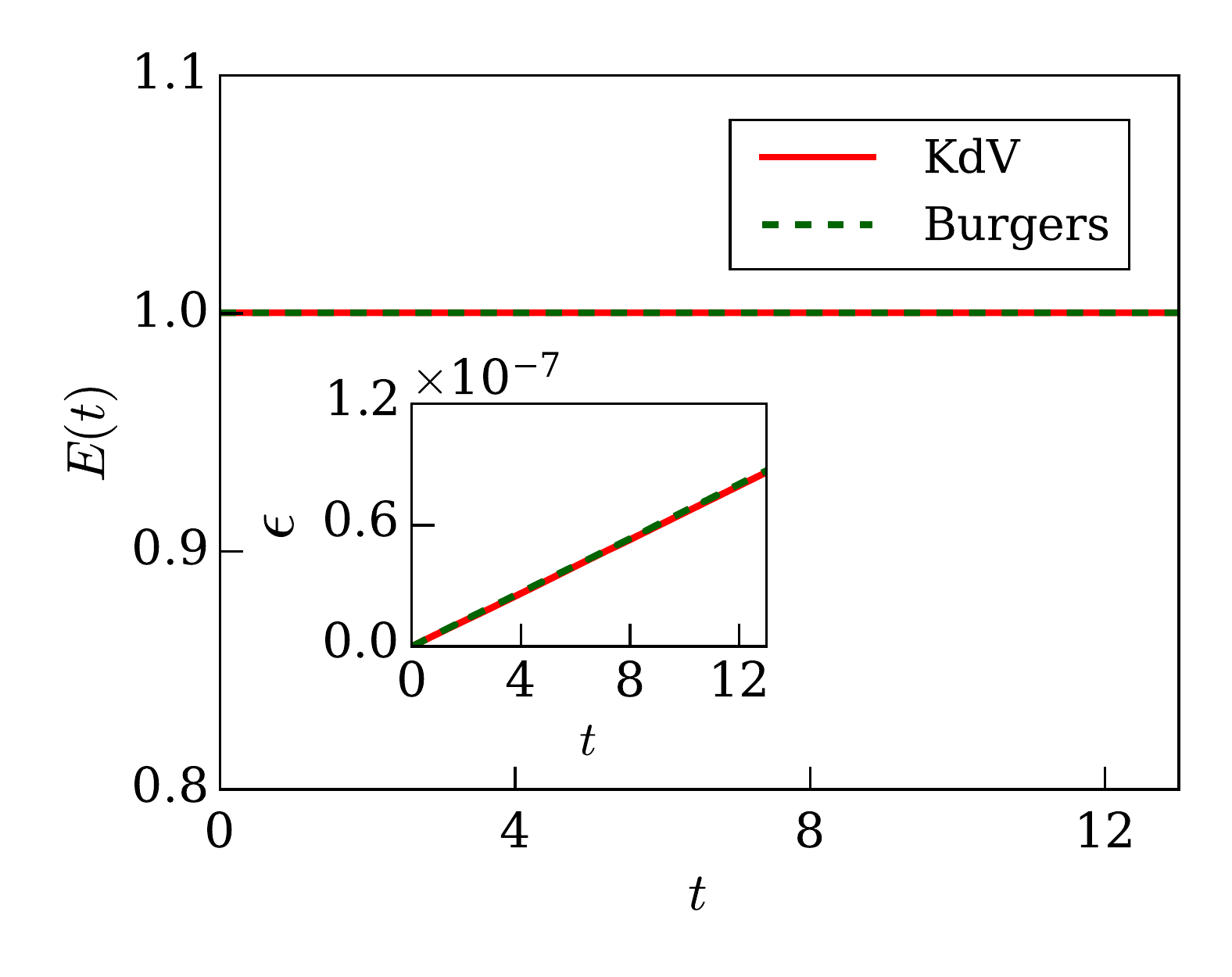}
	\caption{A plot illustrating the constancy of the  total energy for the KdV equation (red curve) and the dissipation-less Burgers equation (green dashed curve). The inset illustrates errors for the two equations.}
	\label{fig:total_energy}
\end{figure}

The temporal evolution of the total energies of the KdV and dissipation-less Burgers equation are shown in Fig.~\ref{fig:total_energy}.  For the whole duration, the relative errors for the two equations are { of th order of $ 10^{-7}$}.   The { RK4 scheme with small $dt$} helps us achieve this accuracy.

\begin{figure}[h!]
	\includegraphics[width=\linewidth]{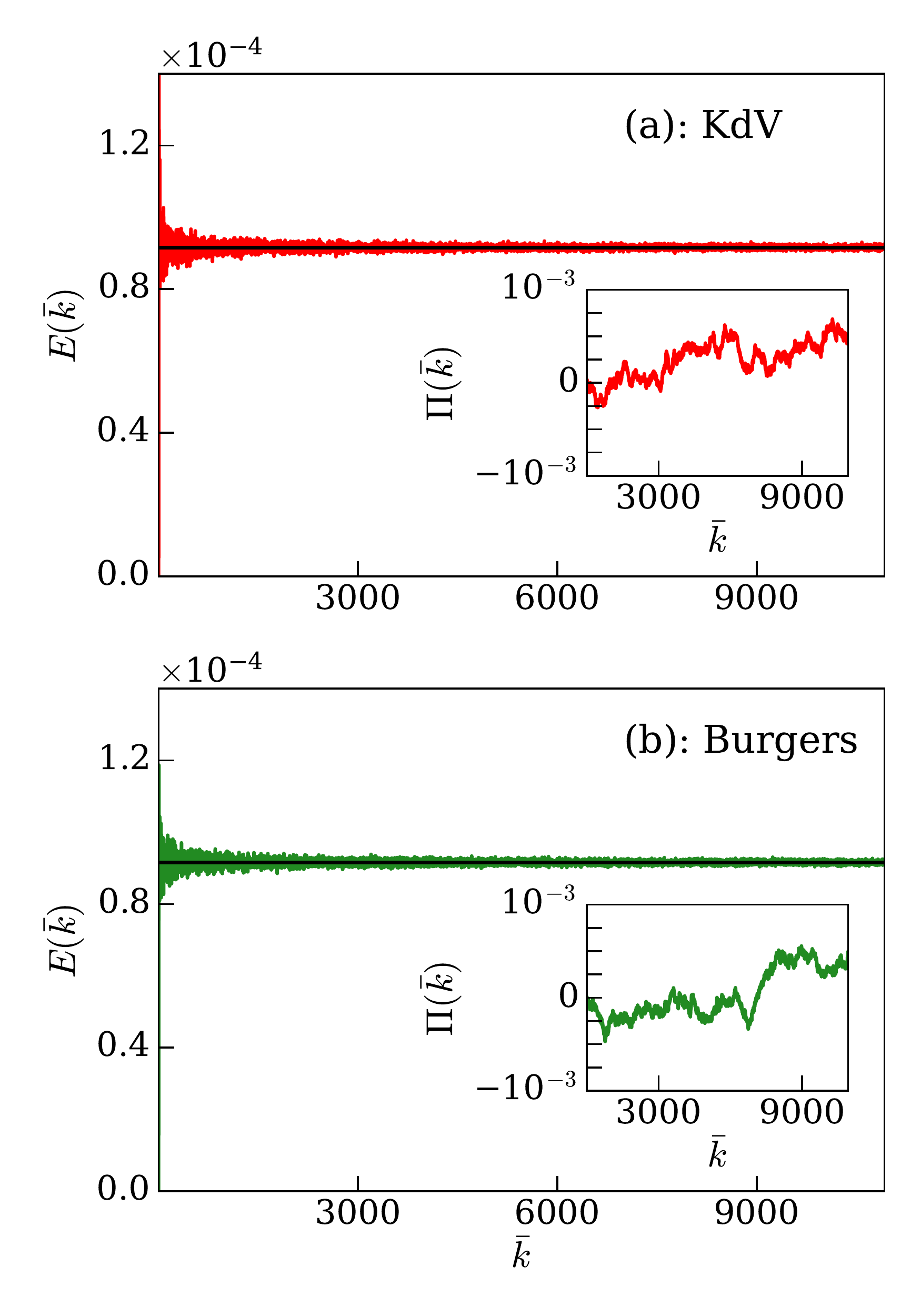}
	\caption{Plots of the   kinetic energy spectra $E(\bar{k}) $  for (a) the KdV equation and (b)  the Burgers turbulence. These spectra are approximately constant with the average values being {$9.2 \times 10^{-5}$ for both the cases}. The average energy fluxes, shown in the insets, aere nearly zeros.} 
	\label{fig:spectra}
\end{figure}


{  
	Next, we  analyze the  energy spectra  of the two equations.  We average these quantities over all the frames ($13 \times 10^6$), and then plot the  energy spectra, $E(\bar{k})$, in Fig.~\ref{fig:spectra}.  We observe that the energy spectra are nearly constant in $k$, with  $E(\bar{k}) \approx 9.2 \times 10^{-5}$ for both the equations. These observations indicate that the energy is 
	equipartitioned among the Fourier modes.    Note that $k_{max}=G/3$, and  that $E(\bar{k})$ is a sum of modal energies of $\hat{u}(k)$ and $\hat{u}(-k)$.  Hence, the average modal energies $ \langle E(k) \rangle $ for the two equations are $4.6 \times 10^{-5}$, which is $ E/N  $.
 
For the two equations, we compute the  energy fluxes, $ \Pi(\bar{k}) $, using Eq.~(\ref{eq:Tk_Pik}).   The averaged fluxes, plotted in the insets of Fig.~\ref{fig:spectra}(a, b), are approximately zeros (less than $10^{-3}$). The vanishing energy flux implies that  there is no net energy transfer (statistically) among the Fourier modes. That is, there is a detailed balance of energy transfers in these systems. These  observations are signatures of equilibrium nature of the systems.

The fluctuations in  $ E(\bar{k}) $ at small wavenumbers are due to an imbalance of energy transfers near $ k=0 $.  This feature appears to be related to the finiteness of the box or to \textit{finite-size scaling}~\cite{Stanley:book:Critical}.   Also, the energy of each Fourier mode  fluctuates around a mean, thus indicating the dynamical nature of the systems.  }

\subsection{PDFs of $ u(x) $ and $ \hat{u}(k) $}

\begin{figure}[h!]
	\includegraphics[width=\linewidth]{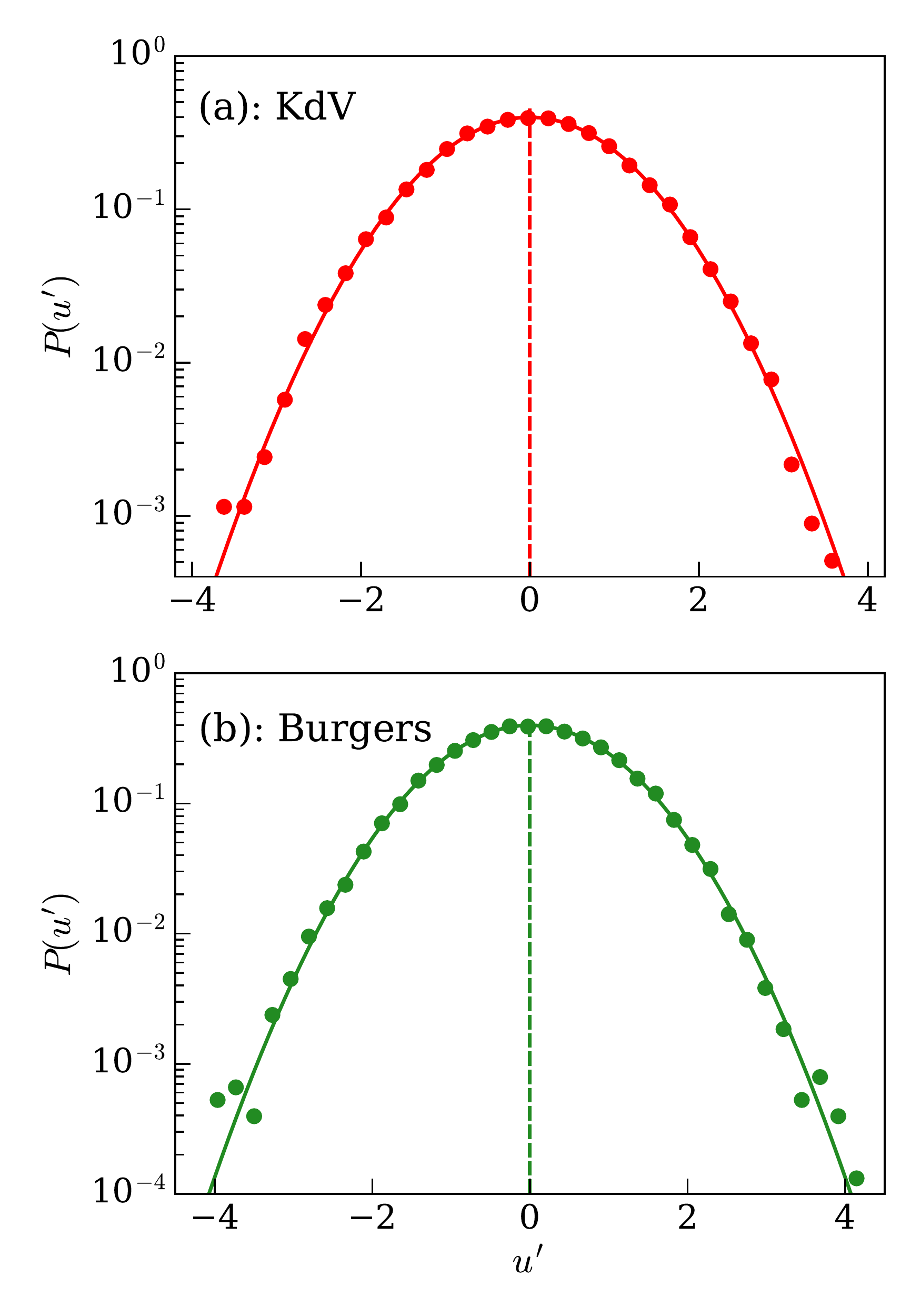}
	\caption{Probability distribution functions $(P(u/u_{\mathrm{rms}}))$ of the normalised real-space velocity magnitude $u/u_{\mathrm{rms}}$ for (a) the KdV equation, and (b) the Burgers equations. Here {$u_{\mathrm{rms}}=1.4$} for the KdV and  Burgers equations. The numerical PDFs exhibit Gaussian distribution.} 
	\label{fig:vel_pdf}
\end{figure}

To verify the equilibrium nature of the field variables, we study the  probability distribution functions (PDF) of $ u(x) $ (real space field) and those of the Fourier modes.   The PDFs of the normalized real-space field $u' = u/u_\mathrm{rms}$ for a single time frame (at {$t=12$}) exhibits  Gaussian distribution (see   Fig.~\ref{fig:vel_pdf}):
\be
P(u')= \frac{1}{\sigma \sqrt{2 \pi}} \exp{\left[-\frac{u'^2} {2 \sigma^2}\right]}
\label{eq:Gaussian_dist}
\ee
with $ \sigma=1 $.  The above PDF demonstrates that the systems are in equilibrium, similar to  a thermodynamic gas.

\begin{figure}[htbp]
	\includegraphics[width=\linewidth]{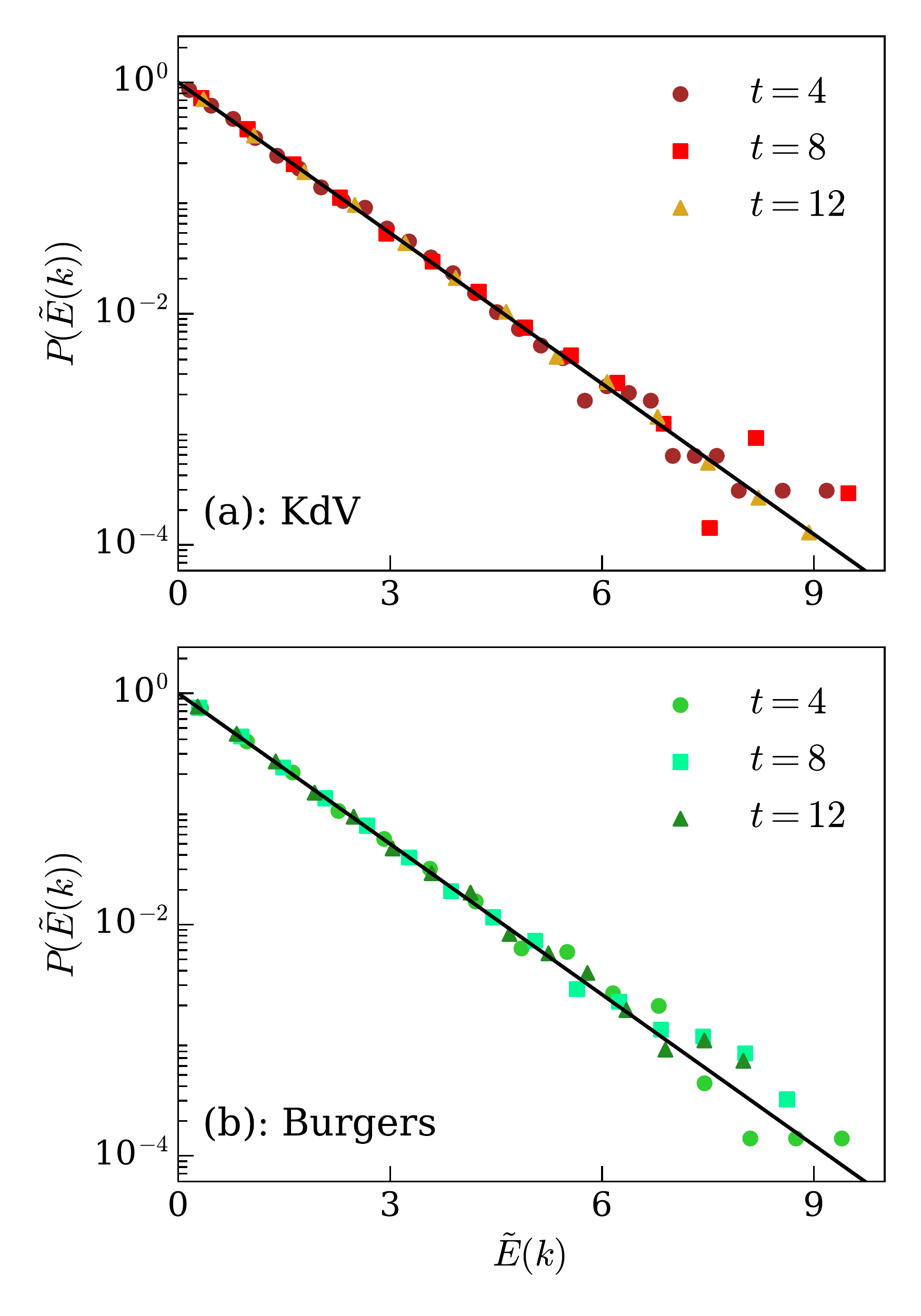}
	\caption{ (a) Probability distribution functions of nondimensionalised modal kinetic energy $\widetilde{E}(k)$ (averaged over all $ k $'s) at {$t=4, 8,$ and $12$} for (a)  the KdV equation, and (b)  the Burgers equation. Here, $ \tilde{E}(k) = E(k) /  \la E(k) \ra $.}
	\label{fig:energy_pdf}
\end{figure}

As discussed in Sec.~\ref{sec:Theory}, the PDF of the modal energy follows Boltzmann-Gibbs distribution:
\be
P(E(k)) = \beta \exp[-\beta E(k)],
\ee
where $\beta$ is  the inverse of the average modal energy ($ E/N  $). For convenience, we plot the PDF of the normalized modal energy, which is $ \tilde{E}(k) = E(k) /  \la E(k) \ra $.  Since $  \la\tilde{E}(k) \ra $ = 1, the PDF of the normalized modal energy $ \tilde{E}(k) $ is
\be
P(  \tilde{E}(k) ) = \exp [ - \tilde{E}(k) ].
\label{eq:Boltzmann}
\ee

In Fig.~\ref{fig:energy_pdf}(a,b) we plot $ P(  \tilde{E}(k) )  $ for the KdV and Burgers equations.  For a better averaging, we compute the PDFs using an ensemble of  $\{ \tilde{E}(k) \}$ at {$ t=4$, 8, and $12$}.  We observe that $ \tilde{E}(k) $ follows Boltzmann distribution   (see Eq.~(\ref{eq:Boltzmann})). Thus, the PDFs of the real space-space velocity field and that of modal energy verify the equilibrium behaviour of the KdV and dissipation-less Burgers turbulence.

%


We may relate the aforementioned randomness to that of a loaded coin. Consider $ N_+$  ($ N_- $) as the count of Fourier modes that have energy more (less) than  the average energy ($ E/N $),   and $ N= N_+ +  N_- $.   We expect that  the following quantity 
\be
\frac {\Delta N}{N} = \frac {N_-}{N}  -  \frac {N_+}{N},
\label{eq:Delta_N}
\ee
to follow a Gaussian distribution. Following the Boltzmann distribution of Eq.~(\ref{eq:Boltzmann}) we  derive that
\bea
\frac{\la N_+ \ra }{N} &=&  \int_{1}^{\infty} e^{- \widetilde{E} } d \widetilde{E} = \frac {1}{e} , \nonumber \\
\frac{\la N_- \ra}{N} &=&  \int_{0}^{1} e^{- \widetilde{E} } d \widetilde{E} = 1-\frac {1}{e} , \nonumber \\
\frac{\la \Delta N \ra}{N}  &=& \frac {\la N_-\ra}{N}  -  \frac { \la N_+\ra}{N}  =1-\frac{2}{e} \approx 0.2642.
\label{eq:delta_N}
\eea
Similarly, we can derive that the standard deviation is 
\be
\sigma_{\Delta N/N} = \frac{1}{\sqrt{N}}\sqrt{\left(1-\left(1-\frac{2}{e}\right)^{2}\right)}.
\ee
The computation of the above quantities and PDF of $ (\Delta N)/N $ requires averaging overa  large number of data.  We observe that the present data is not sufficient for convergence. We plan to compute these quantities in future.



\subsection{Phase space projections of the trajectories}

A major assumption of equilibrium statistical physics is  \textit{ergodic hypothesis}, according to which a system in equilibrium  fills up the available phase space uniformly.  A natural question is whether dissipation-less Burgers and  KdV systems satisfy ergodic hypothesis.  These systems have a enormous degrees of freedom: $ 2N $, with $ N $ each for the real and imaginary parts of $ \hat{u}(k) $.  Hence, it is very difficult to test the ergodic hypothesis directly. However, following \citet{Majda:PNAS2000}, we can argue that the equipartition of energy among the Fourier modes may suggest ergodicity for the two systems.

\begin{figure}[h]
	\includegraphics[width=\linewidth]{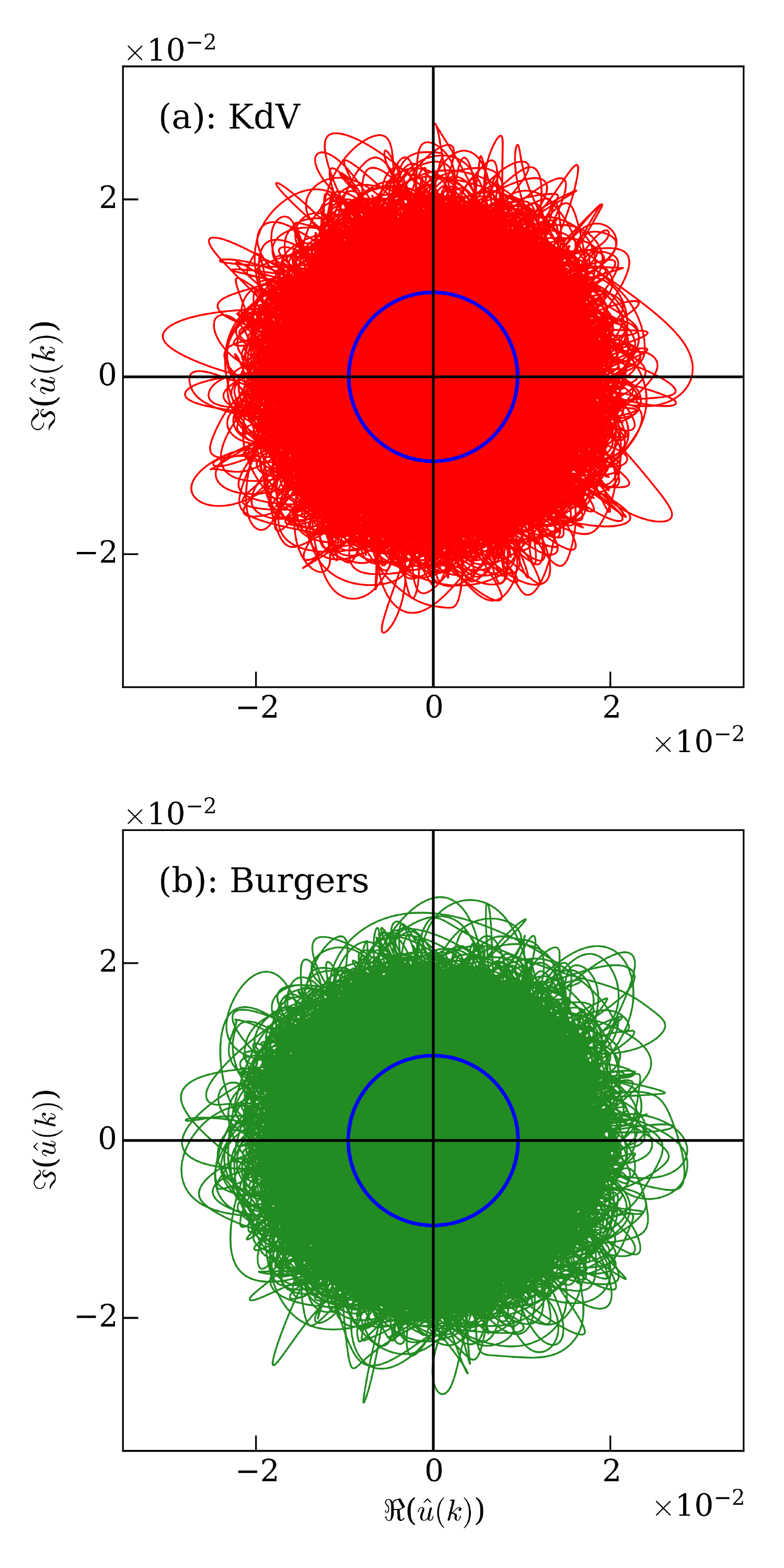}
	\caption{Phase space projections on the 
		$ \{\Re(\hat{u}(k) ) ,  \Im(\hat{u}(k))   \}$ plane with $k=5120$  (a) For the KdV equation;  and (b) for the Burgers equation respectively. The radii of blue circles are corresponding $\sqrt{\la |\hat{u}(k)|^{2} \ra}$.}
	\label{fig:phase_space}
\end{figure}

To investigate the nature of phase space of the KdV and Burgers equations,  we plot the  phase space projections 
$ \{\Re(\hat{u}(k) ) ,  \Im(\hat{u}(k))   \}$ for a generic wavenumber {$k=5120$} (see Fig.~\ref{fig:phase_space}).  For both the equations, we observe that  the trajectories fill up a circular region of the phase space projection nearly uniformly.  Also,  the extreme values of  $ \Re(\hat{u}(k) )$ and  $\Im(\hat{u}(k)) $ are within a factor of $3\sqrt{\la |\hat{u}(k)|^{2} \ra}$ till {$ t=13 $} eddy turnover time.  We expect the fluctuations  to grow with time.  These observations suggest that the phase space trajectory may fill the available phase space, as is conjectured by ergodic hypothesis.

The above observations indicate that the Burgers and KdV equations with delta-correlated noise as initial conditions exhibit equilibrium behaviour.  In the next section, we will discuss another configuration of the KdV equation that exhibits equilibrium behaviour.

\section{Energy spectrum and flux of a soliton  of KdV equation}
\label{sec:Soliton}
		
\begin{figure}[h]
	\includegraphics[width=\linewidth]{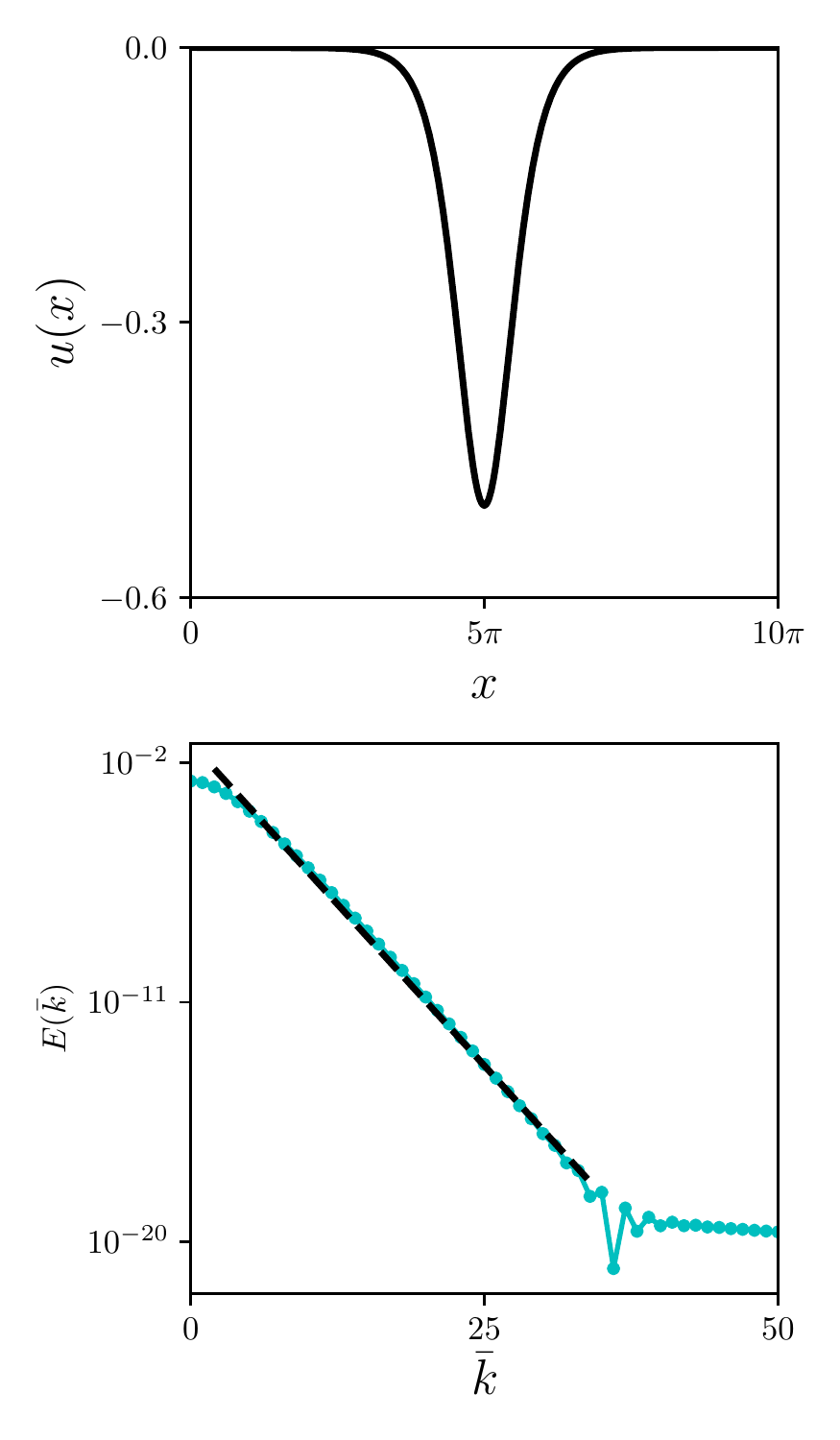}
	\caption{The energy spectrum and flux for the soliton solution of Eq.~(\ref{eq:soliton}). The best-fit curve, $ \exp{[-(k/k_d)] }$, is the black-dashed line.}
	\label{fig:soliton}
\end{figure}

In this section, we show that the soliton solution of KdV equation too is in equilibrium, at least from the energy flux perspective.  To demonstrate the above statement, we work  with another version of KdV equation:
\be
u_t + 6 u u_x + u_{xxx} = 0.
\label{eq:KdV_soliton}
\ee
Equation~(\ref{eq:KdV_soliton}) admits the following soliton  solution~\cite{Drazin:book:Solitons}:
\be
u(x,t) = -\frac{c}{2} \mathrm{sech}^2 
\left[ \frac{\sqrt{c}}{2} (x-ct-x_0) \right]
\label{eq:soliton}
\ee
The above soliton  moves to the right with a constant velocity of $ c $.  { In Fig.~\ref{fig:soliton}(a), we exhibit a snapshot of the soliton for $ c=1 $.

We compute the energy spectrum and flux for the soliton of Eq.~(\ref{eq:soliton}). For these computations, we consider $ u(x) $ to be periodic in a large box of size $ 10 \pi $.  The energy spectrum for the soliton, illustrated in  Fig.~\ref{fig:soliton}(b), is 
\be
E(\bar{k})  = 0.055 \exp{(-1.12 \bar{k})}. 
\ee
The energy flux computed using Eq.~(\ref{eq:Tk_Pik}) is zero (to numerical precision of $ 10^{-19} $).} Vanishing energy flux is consistent with the fact that the KdV equation represents a conservative system.  Hence, for the steady state, following Eq.~(\ref{eq:Fourier_space}), we deduce that
\be 
\la T(k) \ra = 0~~~\mathrm{and}~~~\la \Pi(\bar{k})  \ra = 0.
\ee
  Thus, the  soliton of the KdV equation is  in equilibrium.  This is somewhat surprising because the soliton is  an ordered structure, contrary to the random solutions  discussed in the earlier section.  Also note that the soliton solution is stable due to an absence of energy exchange among the Fourier modes. 

The nature of equilibrium for the soliton and noisy solution of the KdV equation are different.    Since the soliton does not change with time (apart from space translation), its  Fourier modes $ \hat{u}(k)  $  are constant in time. Hence, the soliton is a point in the phase space, and it is not ergodic. Also,  soliton's Fourier modes have different energies (see Fig.~\ref{fig:soliton}), in contrast to an equal distribution of energy among the Fourier modes of the noisy Burgers and KdV equations in equilibrium.  


The KdV equation also admits multiple solitons.  Our preliminary investigation indicates that multiple solitons exhibit nozero energy flux, which is due to the interactions among the constitutive solitons.  A detailed investigation of energy transfers for the multiple solitons will be performed in future.  In addition, it will be interesting to study the evolution of  energy flux for a large-scale initial condition, such as $ \sin(x) $.

	
\section{Discussion and Conclusions} 
\label{sec:conclusions}

In this paper, we showed that for the Burgers and KdV equations, the delta-correlated random initial condition  yields equilibrium or thermalized states.  We observed that these equilibrium states are similar to those of thermodynamic systems.  For example,  the PDF of  the real-space field  is Gaussian, and that of a Fourier mode follows  Boltzmann distribution.

This is an interesting observation considering the fact that the two equations are known for their ordered  structures---shocks for the Burgers equation, and solitons for the KdV equation.  Present simulations show that initial condition plays a critical role in determining the system behaviour.  It is important to contrast the delta-correlated random fields presented in this paper with the mixed states of Burgers equation, which  have been obtained using ordered initial condition, such as $ \sin(x) $ (e.g., see \cite{Ray:PRE2011}).     

As we argue in Sec.~\ref{sec:Theory}, the steady state of a conservative  (non-dissipative) system  has zero energy flux or zero energy transfer statistically. Since such states respect detailed balance, hence they are in equilibrium.   Delta-correlated random initial condition  yields such states quite easily for  the Burgers and KdV equations.   Note however that the mixed state of \citet{Ray:PRE2011} is not steady ($ T(k) \ne 0 $); in this nonequilbrium state, the energy at large scales cascades to small scales, where they thermalize due to the nondissipative nature of the system.  The above mixed state approach equilibrium asymptotically \cite{Ray:PRE2011}.  

Interestingly, a single soliton is a steady solution  of the KdV equation.  Hence, the soliton has no net energy flux, hence it is in equilibrium.  This  important observation  tells us that ordered and non-ergodic structures could be in equilibrium as well; it is not necessary for the equilibrium states to be random~\cite{Majda:PNAS2000}. 

The laws of thermodynamics and  equilibrium statistical mechanics are often applied to systems that appear to be far from equilibrium.  For example, thermodynamics is applied to heat engines, within which flows may be turbulent.  The results of this paper indicate that  random initial condition may be playing an important role in keeping such seemingly nonequilibrium systems close to equilibrium.  This conjecture however needs a closer examination. 

Thus,  one-dimensional Burgers and KdV equations exhibit interesting  equilibrium states.  It will  be interesting to explore  issues such as thermalization,  Poincar\'{e} recurrence theorem, ergodic hypothesis, etc.~for these equations.

\section*{Acknowledgements}
MKV thanks Stephan Fauve, Samriddhi Sankar Ray,  and Arul Laxminarayan for useful discussion.  This work  is partially supported by CEFIPRA project 6104-1 and IFCAM project MA/IFCAM/19/90. Soumyadeep Chatterjee is supported by INSPIRE fellowship (IF180094) from Department of Science \& Technology, India.

\end{document}